\nonstopmode
\documentclass{lmcs} %%% last changed 2014-08-20
\pdfoutput=1

\usepackage{lastpage}
\lmcsdoi{16}{2}{14}
\lmcsheading{}{\pageref{LastPage}}{}{}%
{Feb.~03,~2020}{Jun.~30,~2020}{}


\usepackage{changepage} % adjustwith environment
\usepackage{fleqn}
\usepackage{ifthen}
\newcommand{\citeyearpar}[2][]{\ifthenelse{\equal{#1}{}}{\cite{#2}}{\cite[#1]{#2}}}
\newcommand{\citep}[2][]{\ifthenelse{\equal{#1}{}}{\cite{#2}}{\cite[#1]{#2}}}
\usepackage{xspace}
\usepackage{xcolor}
\usepackage[utf8x]{inputenc}
\usepackage{amssymb}

\usepackage{color}
\definecolor{keywordcolor}{rgb}{0.7, 0.1, 0.1}   % red
\definecolor{commentcolor}{rgb}{0.4, 0.4, 0.4}   % grey
\definecolor{symbolcolor}{rgb}{0.0, 0.1, 0.6}    % blue
\definecolor{sortcolor}{rgb}{0.1, 0.5, 0.1}      % green

\usepackage{listings}

\newcommand{\otto}{\leftrightarrow}
\newcommand{\Type}{\ensuremath{\mathsf{Type}}\xspace}
\newcommand{\Prop}{\ensuremath{\mathsf{Prop}}\xspace}
\newcommand{\funT}[3]{\Pi\,#1:#2.\;#3}
\newcommand{\lam}[3]{\lambda\,#1{:}#2.\;#3}
\newcommand{\tEq}{\_{=}\_}
\newcommand{\Eq}[3]{#2=_{#1}#3}
\newcommand{\tcast}{\ensuremath{\mathsf{cast}}\xspace}
\newcommand{\cast}[4]{\tcast\;#1\;#2\;#3\;#4}
\newcommand{\tEqrec}{\ensuremath{\mathsf{Eq\_rec}}\xspace}
\newcommand{\Eqrec}[6]{\tEqrec\;#1\;#2\;#3\;#4\;#5\;#6}
\newcommand{\tpropext}{\mathsf{propext}}
\newcommand{\ttautext}{\mathsf{tautext}}
\newcommand{\tid}{\mathsf{id}}

\begin{document}

\title[Normalization Fails for Impredicative Proof-Irrelevant Equality]
 {Failure of Normalization in Impredicative Type Theory with
  Proof-Irrelevant Propositional Equality}
% \titlecomment{{\lsuper*}OPTIONAL comment concerning the title, \eg,
%   if a variant or an extended abstract of the paper has appeared elsewhere.}

\author[A.~Abel]{Andreas Abel}
\address{Department for Computer Science and Engineering, Chalmers and
Gothenburg University}
%\email{andreas.abel@cse.gu.se}
\email{\{andreas.abel,thierry.coquand\}@cse.gu.se}
%\thanks{} %optional

\author[T.~Coquand]{Thierry Coquand}
%\address{Department for Computer Science and Engineering, Chalmers and
%Gothenburg University}
%\email{thierry.coquand@cse.gu.se}

%% required for running head on odd and even pages, use suitable
%% abbreviations in case of long titles and many authors:

%%%%%%%%%%%%%%%%%%%%%%%%%%%%%%%%%%%%%%%%%%%%%%%%%%%%%%%%%%%%%%%%%%%%%%%%%%%

%% the abstract has to PRECEDE the command \maketitle:
%% be sure not to issue the \maketitle command twice!

\begin{abstract}
  Normalization fails
  in type theory with an impredicative universe of propositions and a
  proof-irrelevant propositional equality.
  The counterexample to normalization is adapted from Girard's
  counterexample against normalization of System~F
  equipped with a decider for type equality.
  It refutes Werner's normalization conjecture published in LMCS \citep{werner:lmcs08}.
\end{abstract}

\maketitle

\section*{Introduction}
\label{sec:intro}

Type theories with an impredicative universe \Prop of propositions,
such as the Calculus of Constructions \citep{coquandHuet:CoC},
lose the normalization property in the presence of a \emph{proof-irrelevant}
propositional equality $\tEq : \funT A \Type {A \to A \to \Prop}$
with the standard elimination principle.
The loss of normalization is facilitated already by a coercion
function with a reduction rule
\[
\begin{array}{l}
  \tcast : \funT {A\,B} \Prop {\Eq \Prop A B \to A \to B} \\
  \cast A A e x \rhd x
\end{array}
\]
that does not inspect the equality proof $e : \Eq \Prop A A$ but only
checks whether the endpoints are (definitionally) equal.

The failure of normalization refutes a conjecture by Werner
\citeyearpar[Conjecture 3.14]{werner:lmcs08}.
% and strongly diminishes the hope of decidable
% type checking for type theory with an impredicative universe of
% proof irrelevant propositions and a propositional equality that
% eliminates into \Type.
%
Consistency and canonicity is not at stake; thus, the situation is
comparable to type theory with equality reflection
\citep{MartinLoef84,martinLoef:constrMathAndCompPrg}, aka
\emph{Extensional Type Theory}.
At the moment, it is unclear whether the use of impredicativity is essential to
break normalization; predicative type theory might be able to host a
proof-irrelevant propositional equality \citep{abel:nbe09}
while retaining normalization.

\section*{Counterexample to Normalization}

We employ the usual impredicative definition of absurdity $\bot$ and
negation $\neg A$ and a
derived definition of truth $\top$:
\[
\begin{array}{lcl}
  \bot   & : & \Prop \\
  \bot   & = & \funT A \Prop A
\end{array}
\qquad\qquad
\begin{array}{lcl}
  \neg   & : & \Prop \to \Prop \\
  \neg A & = & A \to \bot
\end{array}
\qquad\qquad
\begin{array}{lcl}
  \top   & : & \Prop \\
  \top   & = & \neg \bot \\
\end{array}
\]
The presence of $\tcast$ allows us to define self-application under
the assumption that all propositions all equal.  The self-application
term $\omega$ refutes this assumption.
\[
\begin{array}{lcl}
  \delta & : & \top \\
  \delta\;z & = & z\;\top\;z
%\\[1ex]
\end{array}
\qquad
\begin{array}{lcl}
  \omega & : & \neg \funT {A\,B} \Prop {\Eq \Prop A B} \\
  \omega\;h\;A & = & \tcast\; \top\, A\; {(h\;\top\,A)}\;
    \delta
%   {(\lam z \bot {z\;\top\;z})}
\end{array}
\]
Impredicativity is exploited in $\delta$ when applying $z : \bot$ to type
$\top = \bot \to \bot$ so that it can be applied to $z$ again.  The
type of %the $\lambda$-term
$\delta$
is $\top$ which we cast to $A$ thanks to
the assumption~$h$ that all propositions are equal.

We build a non-normalizing term $\Omega$ by applying $\omega$ to
itself through $\delta$, reminiscent of the shortest diverging term in untyped
$\lambda$-calculus.
\[
\begin{array}{lcl}
  \Omega & : & \neg \funT {A\,B} \Prop {\Eq \Prop A B} \\
  \Omega\;h & = & \delta\;(\omega\;h)
  % \Omega\;h & = & \omega\;h\;\top\;(\omega\;h) \\
\end{array}
\]
Thanks to the reduction rule of $\tcast$, term $\Omega\;h$ reduces to
itself:
\[
\begin{array}{lclcl}
  \Omega\;h
%  & = & \omega\;h\;\top\;(\omega\;h) \\
  & = & \delta\;(\omega\;h)
%\\
  & \rhd & \omega\;h\;\top\;(\omega\;h) \\
  & = & \cast \top \top {(h\;\top\;\top)} \delta \;(\omega\;h)
%\\
  % & = & \cast \top \top {(h\;\top\;\top)} {(\lam z \bot {z\;\top\;z})}\;(\omega\;h) \\
  % & \rhd & (\lam z \bot {z\;\top\;z})\;(\omega\;h) \\
  % & \rhd &  \omega\;h\;\top\;(\omega\;h) \\
  & \rhd & \delta\;(\omega\;h) \\
  & = & \Omega\;h \\
\end{array}
\]
Thus, normalization is lost in the presence of a hypothesis (free
variable) $h$.  As a consequence, normalization that proceeds under
$\lambda$-abstraction can diverge.  This means that equality of open
terms cannot be decided just by normalization.
%  and is likely undecidable.
% As a consequence, type checking is likely undecidable as well.

The counterexample can be implemented in Werner's %LMCS~2008
type theory with
proof-irrelevance \citeyearpar{werner:lmcs08}, refuting the normalization
conjecture (3.14).  We implement $\tcast$ as instance of Werner's
more general equality elimination rule:
\[
\begin{array}{lcl}
  \tEqrec & : &
   \funT A \Type \funT P {A \to \Type}
    \funT {a\,b} A {P\;a \to \Eq A a b \to P\;b}
   \\
  \Eqrec A P a b x e & \rhd & x \qquad \mbox{if } a = b
\\[1ex]
  \cast A B e x & = & \Eqrec \Prop {(\lam a A \Prop)} A B x e
\end{array}
\]
The term $\Omega$ also serves as counterexample to normalization in
the theorem prover Lean \citep{deMoura:cade15},
% \footnote{Microsoft Research, Lean Theorem
%   Prover, \url{https://leanprover.github.io/}.}
version 3.4.2 \citep{lean342}.
\begin{adjustwidth}{\mathindent}{0em}
\begin{lstlisting}
def False := ∀ A : Prop, A
def Not   := λ A, A → False
def True  := Not False

def delta :  True                      := λ z, z True z
def omega :  Not (∀ A B : Prop, A = B) := λ h A, cast (h True A) delta
def Omega :  Not (∀ A B : Prop, A = B) := λ h, delta (omega h)
\end{lstlisting}
% \begin{lstlisting}
% def false' := ∀ A : Prop, A
% def not'   := λ A, A → false'
% def true'  := not' false'

% def omega  :  not' (∀ A B : Prop, A = B)
%            := λ h A, cast (h true' A) (λ z: false', z true' z)

% def Omega  :  not' (∀ A B : Prop, A = B)
%            := λ h, omega h true' (omega h)
% \end{lstlisting}
\end{adjustwidth}
Infinite reduction can now be triggered with the command
\lstinline{#reduce Omega}, which diverges. %times out.

\section*{A Counterexample Using Propositional Extensionality}

The counterexample of the last section used the absurd assumption that
\emph{all} proposition are equal.  The following counterexample
utilizes just the axiom of propositional extensionality, $\tpropext$,
which is a default axiom of Lean.
In fact, the weaker statement $\ttautext$, which states the equality
of \emph{true} propositions, is sufficient.
\[
  \begin{array}{lcl}
\tpropext & : & \funT{A\,B} \Prop {(A \otto B) \to \Eq \Prop A B} \\
\ttautext & : & \funT{A\,B} \Prop {A \to B \to \Eq \Prop A B} \\
  \end{array}
\]
The counterexample uses the standard impredicative definition of
truth,
%$\top = \funT A \Prop {A \to A}$,
\[
  \begin{array}{lcl}
\top & = & \funT A \Prop {A \to A}
  \end{array}
\]
and a cast from
$\top \to \top$ to $A$, which are both tautologies under the
assumption $a : A$.
\[
\begin{array}{lcl}
%\tid   & : & \top \to \top \\
\tid,\; \delta   & : & \top \to \top \\
\tid\;x & = & x \\
% \\[1ex]
% \delta & : & \top \to \top \\
\delta\;z & = & z\;(\top \to \top)\;\tid\;z
%\\[1ex]
\end{array}
\quad
\begin{array}{lcl}
\omega,\; \Omega & : & \top \\
\omega\;A\;a & = &
  \tcast\; (\top \to \top)\; A\;
    (\ttautext\;(\top \to \top)\;A\;\tid\;a)\;
    \delta \\
\Omega & = & \delta\;\omega
% \omega & : & \top \\
% \omega\;A\;a & = &
%   \tcast\; (\top \to \top)\; A\;
% %\\ && \qquad
%     (\ttautext\;(\top \to \top)\;A\;\tid\;a)\;
% %\\ && \qquad
%     (\lam z \top {z\;(\top \to \top)\;\tid\; z})
% \\[1ex]
% \Omega & : & \top \\
% \Omega & = & \omega\;(\top \to \top)\;\tid\;\omega
\end{array}
\]
These definitions can be directly replayed in Lean 3.4.2 with the
standard prelude, yielding a
non-normalizing term \lstinline|Omega|.
\begin{adjustwidth}{\mathindent}{0em}
%  \begin{small}
\begin{lstlisting}
def tautext {A B : Prop} (a : A) (b : B)
          : A = B        := propext (iff.intro (λ _, b) (λ _, a))
def True  : Prop         := ∀ A : Prop, A → A

def delta : True → True := λ z : True, z (True → True) id z
def omega : True         := λ A a, cast (tautext id a) delta
def Omega : True         := delta omega
\end{lstlisting}
%  \end{small}
\end{adjustwidth}
Note that term \lstinline|Omega| is closed with respect the standard axioms
of Lean, and does not even have a weak head normal form.

\section*{Related Work and Conclusions}

The $\tcast$ operator is inspired by Girard's operator
$J : \funT {A\,B} \Prop {A \to B}$ with reduction rule
$J\;A\;A\;M \rhd M$ that destroys the normalization property of
System~F \citep{girard:ssls70,harperMitchell:ipl99}.
In contrast to $J$, our $\tcast$ also requires a proof of
equality of $A$ and $B$, but this proof is not inspected and thus does not
block reduction if it is non-canonical.  Thus, the simple lie that all
propositions are equal is sufficient to trigger divergence.

Historically, the Automath system AUT-4 is maybe the first type-theoretic proof
assistant to feature proof-irrelevant propositions
\citep{deBruijn:automath4}.  The terminology used by de Bruijn is
\emph{fourth degree identification}, where proofs are expressions
considered to have degree 4, propositions and values degree 3, types
and $\Prop$ degree 2, and the universe $\Type$ of types degree 1.

Lean's type theory \citep{carneiro:thesis} features an impredicative
universe of proof-irrelevant propositions which hosts both
propositional equality and the accessibility predicate \citep[1.2]{aczel:inductive75}.  As both may
be eliminated into computational universes, decidability of
definitional equality is lost, as
demonstrated by Carneiro \citeyearpar{carneiro:thesis}
for the case of accessibility.  As a consequence,
typing is not decidable.

%% For equality only (not accessibility):
% Maybe the normalization property for computational
% objects can be salvaged, provided embedded proof terms are not
% reduced, but we have neither proof nor counterexample at the time of writing.

The type-theoretic proof assistants Agda and Coq have recently
\citep{gilbertCockxSozeauTabareau:popl19}
been equipped with a proof-irrelevant universe of propositions
(``strict \Prop'').  In this universe, propositional equality can be
defined, but cannot be eliminated into types that are not strict
propositions themselves.  Under this restriction,
Gilbert \citeyearpar[4.3]{gilbert:PhD}
formally proved normalization and decidability of type
checking for the predicative case.

Several open problems remain:
\begin{enumerate}
\item Does the theory with impredicative strict \Prop have
  normalization and decidability of type checking as well?
\item Does the addition of Werner's rule, while destroying proof
  normalization, preserve decidability of conversion and type
  checking?  (Since proofs
  are irrelevant for equality, they need not be normalized during type
  checking.)
% \item Does the addition of Werner's rule preserve normalization of
%   terms except for the subterms that are proofs?
%   (Our counterexample is a non-normalizing \emph{proof}.)
\item Does Werner's rule preserve normalization in the predicative
  case?  (Our counterexamples make use of impredicativity.)
\end{enumerate}

% \section*{Conclusions}

\subsection*{Acknowledgments}
\noindent
The authors acknowledge support by the Swedish Research Council
(Vetenskapsrådet) under grants 2014-04864 \emph{Termination
  Certificates for Dependently-Typed Programs and Proofs via
  Refinement Types} and
2017-04064 \emph{Syntax and Semantics of Univalent Type Theory}.
Our research group is part of the EU Cost Action
CA15123 The European research network on types for programming and
verification (EUTypes).
We thank Mario Carneiro for contributing the original Lean implementation
of the first counterexample.

\bibliographystyle{alpha}
\bibliography{auto-lmcs20}

\newcommand{\etalchar}[1]{$^{#1}$}
\begin{thebibliography}{dMKA{\etalchar{+}}15}

\bibitem[Abe09]{abel:nbe09}
Andreas Abel.
\newblock Extensional normalization in the logical framework with proof
  irrelevant equality.
\newblock In Olivier Danvy, editor, {\em Workshop on Normalization by
  Evaluation, affiliated to {LiCS} 2009, Los Angeles, 15 August 2009}, 2009.

\bibitem[Acz77]{aczel:inductive75}
Peter Aczel.
\newblock An introduction to inductive definitions.
\newblock In Jon Barwise, editor, {\em Handbook of Mathematical Logic},
  volume~90 of {\em Studies in Logic and the Foundations of Mathematics}, pages
  739--782. Elsevier, 1977.

\bibitem[Car19]{carneiro:thesis}
Mario Carneiro.
\newblock The type theory of {Lean}.
\newblock Master's thesis, Department of Philosophy, Carnegie Mellon
  University, 2019.

\bibitem[CH88]{coquandHuet:CoC}
Thierry Coquand and G{\'e}rard~P. Huet.
\newblock The calculus of constructions.
\newblock {\em Information and Computation}, 76(2/3):95--120, 1988.

\bibitem[dB94]{deBruijn:automath4}
N.G. de~Bruijn.
\newblock Some extensions of {Automath}: The {AUT-4} family.
\newblock In R.P. Nederpelt, J.H. Geuvers, and R.C. de~Vrijer, editors, {\em
  Selected Papers on Automath}, volume 133 of {\em Studies in Logic and the
  Foundations of Mathematics}, pages 283--288. Elsevier, 1994.

\bibitem[dMKA{\etalchar{+}}15]{deMoura:cade15}
Leonardo~Mendon{\c{c}}a de~Moura, Soonho Kong, Jeremy Avigad, Floris van Doorn,
  and Jakob von Raumer.
\newblock The {Lean} theorem prover (system description).
\newblock In Amy~P. Felty and Aart Middeldorp, editors, {\em Automated
  Deduction - {CADE-25} - 25th International Conference on Automated Deduction,
  Berlin, Germany, August 1-7, 2015, Proceedings}, volume 9195 of {\em Lecture
  Notes in Computer Science}, pages 378--388. Springer, 2015.

\bibitem[GCST19]{gilbertCockxSozeauTabareau:popl19}
Ga{\"{e}}tan Gilbert, Jesper Cockx, Matthieu Sozeau, and Nicolas Tabareau.
\newblock Definitional proof-irrelevance without {K}.
\newblock {\em Proceedings of the ACM on Programming Languages},
  3({POPL}):3:1--3:28, 2019.

\bibitem[Gil19]{gilbert:PhD}
Ga\"{e}tan Gilbert.
\newblock {\em A type theory with definitional proof-irrelevance}.
\newblock PhD thesis, \'{E}cole Nationale Sup\'{e}rieure Mines-T\'{e}l\'{e}com
  Atlantique, 2019.

\bibitem[Gir71]{girard:ssls70}
Jean-Yves Girard.
\newblock Une extension de l'interpr\'{e}tation de {G}\"{o}del \`a l'analyse,
  et son application \`a l'\'{e}limination des coupures dans l'analyse et la
  th\'{e}orie des types.
\newblock In J.~E. Fenstad, editor, {\em Proceedings of the {S}econd
  {S}candinavian {L}ogic {S}ymposium ({U}niv. {O}slo, 1970)}, volume~63 of {\em
  Studies in Logic and the Foundations of Mathematics}, pages 63--92. Elsevier,
  1971.

\bibitem[HM99]{harperMitchell:ipl99}
Robert Harper and John~C. Mitchell.
\newblock Parametricity and variants of {Girard}'s \emph{J} operator.
\newblock {\em Information Processing Letters}, 70(1):1--5, 1999.

\bibitem[{Mic}19]{lean342}
{Microsoft Research}.
\newblock Lean theorem prover, 2019.
\newblock Version 3.4.2.

\bibitem[ML84a]{martinLoef:constrMathAndCompPrg}
Per Martin-L\"of.
\newblock Constructive mathematics and computer programming.
\newblock volume 312, pages 501--518. The Royal Society, 1984.

\bibitem[ML84b]{MartinLoef84}
Per Martin-L\"of.
\newblock {\em Intuitionistic Type Theory}.
\newblock Bibliopolis, 1984.

\bibitem[Wer08]{werner:lmcs08}
Benjamin Werner.
\newblock On the strength of proof-irrelevant type theories.
\newblock {\em Logical Methods in Computer Science}, 4(3), 2008.

\end{thebibliography}

\end{document}